\documentclass[12pt]{article}
\usepackage{epsfig}

\parskip=\medskipamount 
plus.3em minus.5em \abovedisplayshortskip=.5em plus.2em minus.4em
\renewcommand{\thesection}{\arabic{section}}

\catcode`@=11

\@addtoreset{equation}{section} \@addtoreset{equation}{subsection}
\def\theequation{\ifnum\value{section}=0 \arabic{equation}\ignorespaces
\else \ifnum\value{section}=-1 A.\arabic{equation}\ignorespaces
\else \ifnum\value{subsection}=0
\thesection.\arabic{equation}\ignorespaces \else
\thesection.\arabic{subsection}.\arabic{equation}\ignorespaces
                             \fi
                        \fi
                   \fi}

{\catcode`\'=\active \def'{{}^\bgroup\prim@s}}

\catcode`@=12

\newcommand{\bq}{\begin{equation}}
\newcommand{\be}{\begin{equation}}
\newcommand{\fq}{\end{equation}}
\newcommand{\ee}{\end{equation}}
\newcommand{\bqr}{\begin{eqnarray}}
\newcommand{\beqs}{\begin{eqnarray}}
\newcommand{\fqr}{\end{eqnarray}}
\newcommand{\eeqs}{\end{eqnarray}}

\newcommand{\rf}[1]{(\ref{#1})}

\def\del{\delta}

\newcommand{\Dslash}{D\!\!\!\!/}

\def\bop#1{\setbox0=\hbox{$#1M$}\mkern1.5mu
    \vbox{\hrule height0pt depth.04\ht0
    \hbox{\vrule width.04\ht0 height.9\ht0 \kern.9\ht0
    \vrule width.04\ht0}\hrule height.04\ht0}\mkern1.5mu}
\def\Box{{\mathpalette\bop{}}}                        

\begin{document}
\thispagestyle{empty}

\vskip .6in
\begin{center}

{\bf N=4 S-duality Compliant Scattering and within the AdS/CFT Duality}

\vskip .6in

{\bf Gordon Chalmers}
\\[5mm]

{e-mail: gordon@quartz.shango.com}

\vskip .5in minus .2in

{\bf Abstract}

\end{center}

It is pointed out the ${\cal N}=4$ supersymmetric gauge theory used in 
the anti-de Sitter holographic correspondence is required to be modified.  
The SL(2,Z) completion of the theory is required to map the correlators 
to string theory amplitudes.  This is overlooked in the literature, and 
requires the (p,q) dyons to be inserted into the gauge theory calculations.   
In addition, the full four-point and higher-point scattering amplitudes 
in the spontaneously broken gauge theory are conjectured, in a similar 
manner as the amplitudes in IIB superstring amplitudes are obtained.  

\vfill\break

\noindent{\it Introduction}

It is understood for years now that there is a holographic duality between 
${\cal N}=4$ supersymmetric SU(N) gauge theory and IIB superstring theory 
compactified in a stable de Sitter background \cite{AdS1}-\cite{AdS3}.  
This duality has been explored in many works; as far as the correlation 
functions are concerned primarily only the three-point functions and 
four-point functions in the supergravity side have been computed due to 
the difficulty in obtaining non-perturbative four-point functions.  The 
microscopic ${\cal N}=4$ 
supersymmetric gauge theory is to possess a non-perturbative S-duality 
that is compatible with the corresponding one in IIB superstring theory.  
The latter is not manifest in the direct quantization of the classical 
Lagrangian.

In this note it is commented on how to complete the fundamental Lagrangian 
in the quantum regime in order to restore S-duality.  For several 
reasons, plus the form and coefficients of the tree amplitudes, a 
completion of the ${\cal N}=4$ scattering is required.  The S-dual 
compliant scattering should be used in the AdS/CFT correspondence.  
Typically, only the perturbative form of the gauge theory is used, 
without the (p,q) dyons in tests of the correspondence.  

\vskip .2in 
\noindent{\it Modular Invariant Derivative Expansion}  

The ${\cal N}=4$ spontaneously broken theory is examined in this work.  
The Lagrangian is, 

\bqr  
{\cal S}={1\over g^2} {\rm Tr}~ \int ~ d^4x \bigl[ F^2 + \phi \Box \phi + 
  \psi {\Dslash} \psi + \left[ \phi,\phi\right]^2 \bigr] \ . 
\fqr 
The quantum theory is believed to have a full S-duality, which means that 
the gauge amplitudes are invariant: under $A\rightarrow A_D$ and 
$\tau\rightarrow (a\tau+b)/ (c\tau+d)$ the functional form of the amplitude 
is invariant.  The series supports a tower of dyonic muliplets satisfying 
the mass formula $m^2=2\vert n^i a_i +m^i a_{d,i}\vert^2$ 
with $a_i$ and $a_{d,i}$ the vacuum values of the scalars and their 
duals; $a_{d,i}=\tau a_i$.  The two couplings parameterizing the simplest 
SU(2)$\rightarrow$U(1) theory is, 

\bqr  
{\theta\over 2\pi} + {4\pi i\over g^2}=\tau=\tau_1+i\tau_2 \ , 
\fqr 
taking values in the Teichmuller space of the keyhole region in the upper 
half plane, i.e. $\vert\tau\vert\geq 1/2$ and $\vert \tau_1\vert \leq 1/2$.  
The S-duality invariant scattering within the derivative expansion is 
constructed in \cite{Chalmers5}.  Derivative expansions in general are 
examined in \cite{Chalmers5}-\cite{Chalmers11}.  

The full amplitudes of ${\cal N}=4$ theory may be constructed either in 
a gauge coupling perturbative series, i.e. the usual diagrammatic expansion 
formulated via unitarity methods, or as an expansion in derivatives, with 
the latter approach being nonperturbative in coupling.  Both expansions are 
equivalent, found from a diagram by diagram basis.  

The full set of operators to create a spontaneously broken ${\cal N}=4$ 
gauge theory amplitude is found from 

\bqr 
{\cal O}= \prod_{j=1} {\rm Tr} F^k_j \ , 
\fqr 
with possible $\ln^{m_1}(\Box) \ldots \ln^{m_n})$ (from the massless 
sector) and combinations with the covariant derivative; the derivatives 
are gauge covariantized and the tensor contractions are implied.  The 
dimensionality of the operator is compensated by a factor of the 
vacuum expectation value, $\langle\phi^2\rangle^m$.  The generic tensor 
includes the fermions and scalars as in.  

The generating function of the gauge theory ${\cal N}=4$ four-point 
amplitude is given 
 
\bqr 
{\cal S}_4 =\sum ~ \int d^dx~ h_n(\tau,\bar\tau) {\cal O}_n \ , 
\fqr 
with the ring of functions spanning $h_n(\tau,\bar\tau)$ consisting of 
the elements, 

\bqr 
g^{n-2} (g^2)^s~ \prod E_{s_j}^{(q_j,-q_j)} (\tau,\bar\tau) \ , 
\fqr 
and their weights  

\bqr 
\sum_j s_j = n/2 \ , \qquad \sum_j q_j = 0 \ , 
\fqr 
The general covariant term in the effective theory has terms,  

\bqr  
\prod_j \nabla_{\rho(i)} 
\prod_{i=1}^{n_\partial} F_{\mu(i)\nu(i)} \prod_{j=1}^{n_i^\phi} 
 \phi_{a_{\rho(j)}} \prod^{m_i^\psi} \psi_{a_{\kappa(j)}}  \ , 
\fqr 
with the derivatives placed in various orderings (multiplying fields and 
products of combinations of fields; this is described in momentum space in 
\cite{Chalmers1}).  The multiplying Eisenstein series possessing weights, 

\bqr  
s=3/16 ~{\rm dim} {\cal O} \qquad q=n_\psi/2 \ . 
\fqr 
These terms span the general operator ${\cal O}$ in the generating 
functional.  The non-holomorphic weight $q$ is correlated with the 
R-symmetry.  

The perturbative coupling structure, for the gauge bosons as an example, 
has the form, 

\bqr 
g^{n-2} (g^2)^{n_{\rm max}/2} \Bigl[ \bigl({1\over g^2}\bigr)^{n_{\rm max}/2} , 
\ldots , \bigl({1\over g^2}\bigr)^{-n_{\rm max}+1} \Bigr] \ . 
\label{couplingexp}
\fqr 
The factor in brackets agrees with the modular expansion of the 
Eisenstein series pertinent to the scattering amplitudes, and the prefactor 
may be absorbed by a field redefinition, 

\bqr 
A\rightarrow {1\over g} A \qquad d^4x ~{\cal O}\rightarrow  
 d^4x~ g^{-n_{\rm max}/2+2} {\cal O}\ , 
\fqr 
which maps the gauge field part of the Lagrangian into 

\bqr 
\int d^4x ~ {1\over g^2} {\rm Tr}\left( \partial A + 
{1\over g} A^2\right)^2 \ ,  
\fqr 
together with a 'Weyl' rescaling of the composite operator (e.g. 
${\rm Tr} F^4$).  
This field redefinition, together with the supersymmetric completion, 
agrees with the ${\cal N}=4$ S-duality self-mapping in a manifest way.  

Fermionic (and mixed) amplitudes would have a non-vanishing $q_j$ sum.  
The Eisenstein functions have the representation 

\bqr 
E_{s_j}^{(q_j,-q_j)} (\tau,\bar\tau) = \sum_{(p,q)\neq (0,0)}  {\tau_2^s\over 
 (p+q\tau)^{s-q} (p+q\bar\tau)^{s+q}} \ , 
\fqr   
with an expansion containing two monomial terms and an infinite number 
of exponential (representing instanton) terms, 

\bqr 
E_s(\tau,\bar\tau) = 2\zeta(2s) \tau_2^s + {\sqrt\pi} {\Gamma(s-1/2)\over 
\Gamma(s)} \zeta(2s-1) \tau_2^{1-2s} + {\cal O}(e^{-2\pi\tau}) \ldots 
\fqr 
with a modification in the non-holomorphic counterpart, $E_s^{(q,-q)}$, 
but with the same zeta function factors.  The latter terms correspond to 
gauge theory instanton contributions to the amplitude; via S-duality all of 
the instantonic terms are available from the perturbative sector.  (At 
$s=0$ or $s={1\over 2}$ the expansion is finite: $\zeta(0)=-1$ and both 
$\zeta(2s-1)\vert_{s=0}$ and $\Gamma(s)\vert_{s=0}$ have simple poles.)  
The $n$-point amplitudes, with the previously discussed modular weight, are 

\bqr 
\langle A(k_1) \ldots A(k_n)\rangle = \sum_q h_q^{(n)}(\tau,\bar\tau) 
f_q(k_1,\ldots,  k_n) \ , 
\fqr 
where the modular factor is h (with the weights $n_A/2+2$) and the 
kinematic structure of the higher derivative term $f_q$.  

The sewing relations that allow for a determination of the coefficients 
of the modular functions at the various derivative orders is not reviewed.  
This is discussed in detail in \cite{Chalmers5}-\cite{Chalmers7}.  

In \cite{Chalmers11} the derivative expansion in IIB superstring theory 
was formulated also in terms of the appropriate modular invariant forms.  
Due to a conjecture of Green \cite{Green1} the linear combinations of 
these forms was delimited, to all orders in \cite{Green1}-\cite{Chalmers12}.  
In simple number theory terms, this conjecture was further established 
and shown to follow from the partitions of the deformed free boson, 

\bqr 
P=\prod {1\over 1-2x^{2n+1}} \ .  
\label{partition}
\fqr 
The simplicity of the partitions of \rf{partition} does appear to 
suggest the correctness of the modular form ansatz.  Also, the partitions 
suggest an affine symmetry of the scattering.  

\vskip .2in 
\noindent{\it IIB Scattering and Number Theoretic Description}  

The modular ansatz for the IIB superstring scattering is generated by  
replacing the zeta functions in the (4-pt) tree amplitude, 

\bqr  
\prod_j \zeta(2p_j+1) \rightarrow Z_{\left\{p_j+1/2\right\}} \ , 
\fqr 
with the $Z$ functions described in the next section. 

The functions $Z_{\left\{ p_j+1/2\right\} }$ are described by the modular 
invariant differential equation on the torus, 

\bqr 
{1\over 4} \Delta Z_{\left\{ q_j \right\} } = 
 A Z_{ \left\{ q_j \right\} } + B \prod_j Z_{q_j} \ , 
\fqr 
with the simplest case being the Eisenstein functions, 

\bqr 
Z_s=E_s  \qquad s(s-1) Z_s = \Delta Z_s \ . 
\fqr 
The Laplacian takes the form, when restricted to the perturbative sector, 
that is, without the $\tau_1$ dependence, 

\bqr 
\Delta=4\tau_2^2\del_\tau\del_{\bar\tau} \ .  
\fqr 
The condition on $A$ and $B$ could in principle be determined generically 
by the tree and one-loop contributions of the usual perturbative string 
amplitude; however, their numbers are left unknown for the moment.
 
There have been several proposals for the quantum completion of the 
S-matrix, and higher derivative terms up to genus two have been computed.  
The modular invariant completion due to S-duality enforces certain 
structures on the coupling dependence.  A basis for the coupling structure 
is formed from the Eisenstein functions, the contribution of which have 
recently been elucidated more completely in \cite{Green1}.  

The polynomial system generating the perturbative contributions 
can be determined from a graphical illustration and also through 
a 'vertex' algebra.  The latter can be found from expanding the 
function, 

\bqr 
\prod_{n=1} {1\over (1-2 x^{2n+1})} \ , 
\label{modularexp}
\fqr   
which is similar to the partition function of a boson on the 
torus, 

\bqr 
\prod_{n=0} {1\over (1-x^{2n+1})} \ .  
\fqr 
The latter is associated to a vertex algebra.  The former 
will be shown to correspond to the perturbative four-point function, 
without the non-analytic terms required by unitarity.

There are a set of trees as depicted in figure 1.  Each tree is 
found by taking a number $N$, an odd number, and partitioning it 
into odd numbers 
$3,5,7,\ldots$.  At each pair of nodes the numbers in the partition 
are attached.  The partition is labeled by the set $N(\{\zeta\})$.  

\begin{figure}
\begin{center}
\epsfxsize=12cm
\epsfysize=12cm
\epsfbox{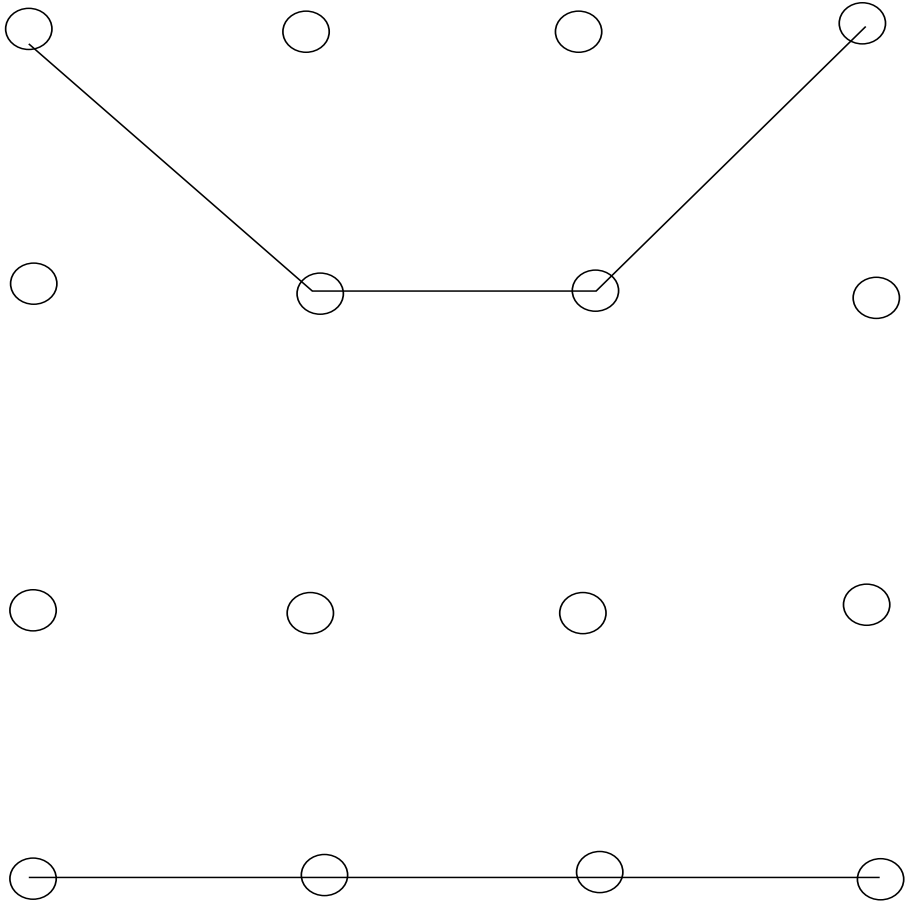}
\end{center}
\caption{Weighted trees.}
\end{figure}

The nodes of the trees are chosen, one from each pair, and a set 
of lines could be drawn between these nodes.  For each tree there are  

\bqr 
{a!\over b!(b-a)!}  
\label{treenumber}
\fqr 
ways of partitioning the tree into the same number of 'up' and 'down' 
nodes.  Each tree set is labeled by $a=N(\{\zeta\})$, and the terms 
in \rf{treenumber} are spanned by $b=0,...,N(\{\zeta\})$ for 
$a=N(\{\zeta\})$.  There are $2^{N(\{\zeta\})}$ terms or polynomials 
in each tree, as found by summing \rf{treenumber}.     

The derivative terms are labeled by $\Box^{2k}R^4$ with $k$ either 
half integral or integral starting at $k=0$.  At a specific order 
$\Box^{2k}$, the tree system is found by partitioning the number 
$N=2k+3$.  The number of partitions a number $N$ can have into odd 
numbers, excluding unity, is denoted $P_{\rm odd}(N)$.  The perturbative 
contributions to this derivative order are found as follows, 

1) attach weights to the nodes of the tree 

2) take the product of the weights of the nodes in the tree 

3) add the sums of the products 

\noindent The sum of the products reintroduces the contribution of the 
modular ansatz to the four-point scattering.  
  
For reference, the genus truncation property holds at $g_{\rm max}= 
{1\over 2}(2k+1), {1\over 2}(2k+2)$ for $k=n/2$ with $n$ either odd or even.  
There are $P_{\rm odd}(2k+3)$ partitions of $2k+3$ into odd numbers excluding 
unity, with a maximum number.  The genus truncation follows from the 
fact that there a maximum number of nodes in the tree.

Some examples of the partitions of numbers at a given derivative order 
$\Box^{2k} R^4$ are described in the following table.  For example, at 
$k=4$, the number $11$ can be partitioned into $11$ and $3+3+5$, with 
counts of $2^1$ and $2^3$.  

\bqr   
\pmatrix{ 
      k=0    &  2         &     k=4     & 2+2^3=10 \cr 
      k=1    &  2         &     k=5     & 2+2^3+2^3=18   \cr 
      k=2    &  2         &     k=6     & 2+2^5+2^3+2^3+2^3=68 \cr 
      k=3    &  2+2^3=10  &     k=7     & 2+2^3+2^5+2^3+2^3=58 \cr 
}
\fqr     
One would like to represent the terms group theoretically with the  
quantum numbers being zeta entries.  This can be done using the 
expansion in \rf{modularexp}.  

The perturbative contribution to the order $\Box^{2k}$ can be 
read off of the tree by associating the weights to the nodes.  
For the 'up' node there is a factor 

\bqr 
{2\zeta(2s)\over [(3/2+k)(3/2+k-1)-A]}, 
\fqr 
and for the 'down' node there is 

\bqr 
{{2\sqrt\pi\Gamma(s-1/2)\zeta(2s-1)\over\Gamma(s) 
[(3/2+k-2{\rm gmax})(3/2+k-2{\rm gmax}-1)-A]}} \ .  
\fqr 
Each perturbative contribution is found by multiplying the 
node contributions.  There are the various weighted trees that 
contribute to the perturbative contribution at $2k+3$, when 
partitioned into the various odd numbers.  Each contribution 
has the weighted factor of $B$ in the product.  (The weighted trees 
resemble a fermionic system with gmax fermions with a quantum level 
degeneracy non-identical fermions.)  The partitioning of the 
number $2k+3$ into the weighted trees is a convenient way of 
describing all of the contributions to the particular derivative 
term.

\vskip .2in 
\noindent{\it Analog Number Theoretic Description to IIB} 

In this section a similar conjecture for the modular forms, with 
the appropriate number theoretic interpretation, is generated for 
the ${\cal N}=4$ supersymmetric theory.  The ansatz defines 
specifically the linear combination of modular forms entering 
into the modular ansatz.  

Recall that a specific term in the derivative expansion is paramterized 
by the combinations 

\bqr 
\prod E_{s_i}(\tau,\bar\tau) \qquad \sum s_i = s=n/2 \ .  
\fqr 
The modular ansatz, as presented in the IIB superstring delimits 
the specific linear combination.  

A conjecture for the ${\cal N}=4$ supersymmetric gauge theory 
is that the modular function entering into the $\Box^k F^m$ term, is 
exactly the same as the $\Box^k R^m$ (with $R$ the Ricci curvature).  
For example, the $F^4$ term maps to the $R^4$ term, with the $E_{3/2}$ 
function as a prefactor.  The correspondence holds for $n$-point 
scattering amplitudes.  

Although, as in the IIB superstring scattering, this conjecture 
can in not proven in this work, the maximal supersymmetry suggests 
this.  The partition 

\bqr 
\prod_{n=1} {1\over (1-2 x^{2n+1})} \ , 
\label{modularexp2}
\fqr   
generates the four-point gauge boson scattering amplitude.  The 
integrability of the gauge theory and the type IIB superstring 
has the same origin.  

The supersymmetric completion of the $n$-point scattering in 
the $\Box^k F^m$ generates the fermionic and scalar components, 
in the same manner that the supersymmetric completion of the 
$\Box^k R^m$ terms generate the IIB effective action. 

\vskip .2in 
\noindent{\it AdS/CFT}  

The anti-de Sitter correspondence between ${\cal N}=4$ supersymmetric 
gauge theory and IIB superstring theory requires that the composite 
correlations 

\bqr 
\langle \prod {\cal O}_i (x_i)\rangle 
\fqr 
be modular invariant.  The diagrams in figure 2 depict the gauge 
theory correlations.  

\begin{figure}
\begin{center}
\epsfxsize=12cm
\epsfysize=6cm
\epsfbox{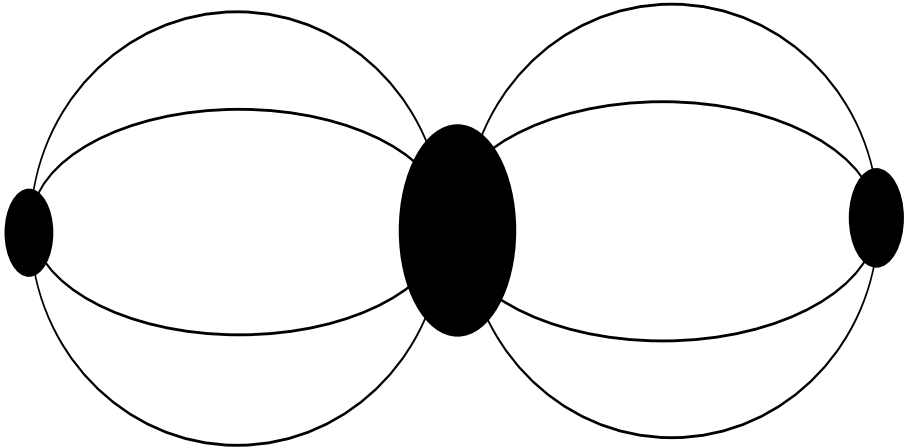}
\end{center}
\caption{Correlation in the ${\cal N}=4$ gauge theory.  The internal 
vertex represents the full four-point scattering in the effective 
action.}
\end{figure}

Due to the microscopic theory, S-duality will only be present 
in the correlation functions if the (p,q) dyonic states appear 
in the effective action (and scattering).  A scattering amplitude 
built from the perturbative gauge theory, and instantons, will 
not by S-duality compliant.   The most direct way to see this 
is that the tree amplitudes in the S-duality compliant scattering 
(i.e. the tree contributions to $\Box^k F^m$) must contain factors 
of the zeta function.  The modular construction predicts these 
zeta factors appearing after expanding the Eisenstein series.  
The perturbative gauge theory with instantons cannot produce these 
zeta values; rather the (p,q) dyonic states must be included.  
(It appears unnatural to multiply the Eisenstein series by 
additional zeta functions, because the loop contributions would 
contain odd ratios of zeta functions).  

The modular invariant scattering between the IIB superstring 
and the ${\cal N}=4$ gauge theory is naturally described in the 
S-duality context with the conjectured ${\cal N}=4$ scattering 
in this work.   

It should be noted that one purpose of the conjectured holography 
is to compute non-perturbative terms in the gauge theory from 
supergravity and string theory.  The conjectured S-duality compliant 
scattering in the gauge theory, with the $\Box^k F^m$ mapping to 
the $\Box^k R^m$, is a more direct way to obtain the nonperturbative 
gauge theory scattering.  

Although the gauge and string theory scattering is only a conjecture, 
there is evidence for its correctness in the presence of the dyonic 
states, the modular invariance, maximal supersymmetry, and the 
partition function indicating integrability. 

\vskip .2in 
\noindent{\it Conclusion}   

A conjecture for the modular invariant scattering in the ${\cal N}=4$ 
supersymmetric gauge theory has been produced.  This is for each 
operator in the supersymmetric action, and to all orders in the coupling, 
including the instantons.  The conjecture is similar to the recent 
papers describing the supersymmetric action of the IIB superstring, 
which is also reviewed.  The integrability is obvious in these constructions, 
both in the IIB superstring and the ${\cal N}=4$ gauge theory, although 
further diagrammatic work is required to rigourously prove the constructions.  

The modular invariance of the scattering is required to make the correlation 
functions of composite operators in the gauge theory S-dual compliant.  This 
latter property is required both for the quantum action of the gauge 
theory and for the anti-de Sitter holographic correspondence with the IIB 
superstring.  Typically this modular invariance is not invoked in the 
holographic duality due in part because the non-perturbative correlation 
functions havent been computed.  

It would be interesting to find constructions analogous to those presented 
here for theories with lower amounts of supersymmetries.  

\vfill\break

\end{document}